\documentclass[12pt]{article}
\usepackage{amssymb,tabularx,multirow,amsmath,natbib, epsfig, threeparttable, amstext, subfigure}
\usepackage{macros_scharf}
\usepackage{hyperref}
\usepackage{float}
\DeclareGraphicsExtensions{.eps}
\graphicspath{{arXiv_figs/}}

\newcommand{\blind}{0}

\begin{document}

\def\spacingset#1{\renewcommand{\baselinestretch}%
{#1}\small\normalsize} \spacingset{1}


\date{}

\if0\blind { \title{\bf Imputation Approaches for Animal Movement Modeling} \author{Henry Scharf\thanks{\tiny Corresponding author: henry.scharf@colostate.edu}\hspace{.2cm}\\
    Department of Statistics \\ Colorado State University\\
    and \\
    Mevin B. Hooten \hspace{.2cm}\\
    U.S. Geological Survey \\ Colorado Cooperative Fish and Wildlife Research Unit \\ Department of Fish, Wildlife, and Conservation Biology \\ Department of Statistics \\ Colorado State University\\
    and \\
    Devin S. Johnson \\
    Alaska Fisheries Science Center \\ NOAA Fisheries}
  \maketitle
} \fi

\if1\blind
{
  \bigskip
  \bigskip
  \bigskip
  \begin{center}
    {\LARGE\bf Imputation Approaches for Animal Movement Modeling}
\end{center}
  \medskip
} \fi

\bigskip
\pagebreak

\begin{abstract}

The analysis of telemetry data is common in animal ecological studies. While the collection of telemetry data for individual animals has improved dramatically, the methods to properly account for inherent uncertainties (e.g., measurement error, dependence, barriers to movement) have lagged behind. Still, many new statistical approaches have been developed to infer unknown quantities affecting animal movement or predict movement based on telemetry data. Hierarchical statistical models are useful to account for some of the aforementioned uncertainties, as well as provide population-level inference, but they often come with an increased computational burden. For certain types of statistical models, it is straightforward to provide inference if the latent true animal trajectory is known, but challenging otherwise. In these cases, approaches related to multiple imputation have been employed to account for the uncertainty associated with our knowledge of the latent trajectory. Despite the increasing use of imputation approaches for modeling animal movement, the general sensitivity and accuracy of these methods have not been explored in detail. We provide an introduction to animal movement modeling and describe how imputation approaches may be helpful for certain types of models. We also assess the performance of imputation approaches in a simulation study. Our simulation study suggests that inference for model parameters directly related to the location of an individual may be more accurate than inference for parameters associated with higher-order processes such as velocity or acceleration. Finally, we apply these methods to analyze a telemetry data set involving northern fur seals (\textit{Callorhinus ursinus}) in the Bering Sea.

\end{abstract}

\noindent%
\vfill

\newpage

\section{Introduction}
An increasing diversity of telemetry data types and their use in a variety of scientific studies has resulted in rapid development of models for animal trajectories \citep{McClintock2014, Kays2015}. Statistical models for animal movement processes mostly assume one of three forms:  point process models, discrete-time models, and continuous-time models \citep{Hooten2017}. Each of these three types of statistical models are aimed at answering certain characteristic questions. For example, point process models are often used to infer individual- and/or population-level ``resource selection,'' \citep[e.g.,][]{Brost2015}, whereas discrete-time models are commonly used to temporally cluster trajectories into discrete components that may relate to animal behavior \citep[e.g.,][]{Langrock2012, McClintock2013}. Continuous-time models for animal movement have linked trajectories to mechanistic stochastic processes \citep[e.g.,][]{Brillinger2010, Hooten2016} and are also used to interpolate telemetry data; that is, to predict an individual's position at times of interest \citep{Johnson2008, Hooten2017}.

Hierarchical approaches to model specification facilitate population-level inference \citep{Hooten2016a} and account for telemetry measurement error \citep[e.g.,][]{Brost2015, Buderman2016}. Some forms of telemetry measurement error have unique features due to the data collection procedure (e.g., the satellite orbit of Service Argos; \citealt{McClintock2015}) that must be explicitly accounted for in statistical models for animal movement. Hierarchical models provide a rigorous framework that both allows for feedback between the data- and process-level components of a complete movement model, and properly accounts for uncertainty due to measurement error.

While the advantages of hierarchical models are clear, they can often be challenging to fit to real data sets due to their inherent complexities. In some cases, a full hierarchical model may be impossible to implement when the process is specified for a functional of the true trajectory that is insufficient (i.e., carries less information content than the position process). If an animal movement model is specified for a certain type of latent process that does not reconcile well with the telemetry data, then traditional hierarchical modeling approaches may not be feasible. For example, the discrete-space models for movement in \cite{Hooten2010}, \cite{Hanks2015}, and \cite{Hanks2016} involve a transformation of the continuous underlying path into residence times associated with a finite set of areal regions. It is straightforward to calculate the areal residence times based on the true trajectory, but impossible to recover the true trajectory based on the areal residence times alone. This noninvertible functional of the trajectory precludes the ability to fit the full hierarchical model using standard methods (e.g., maximum likelihood or Bayesian algorithms; \citealt{Hooten2010}).

In other settings, it may be possible to specify and fit a fully hierarchical model, but doing so would involve excessive computational resources \citep[e.g.,][]{Scharf2016}. For example, a transformation of the trajectory into one or more of its temporal derivatives results in a consistent representation of the trajectory (if the initial conditions are known); however, if the temporal derivative itself is treated as a latent stochastic process in a hierarchical model, we may not be able to analytically marginalize over the true trajectory to obtain an integrated likelihood. Thus, an algorithm to fit the model, such as Markov chain Monte Carlo (MCMC), must be constructed to sample the latent continuous-time trajectory in addition to the other model parameters. It may not be feasible to quickly obtain samples of the true latent trajectory process in such a model, thus, alternative approaches based on imputation have been considered.

\begin{figure}[H]
  \centering
  \includegraphics[width = 0.67\textwidth]{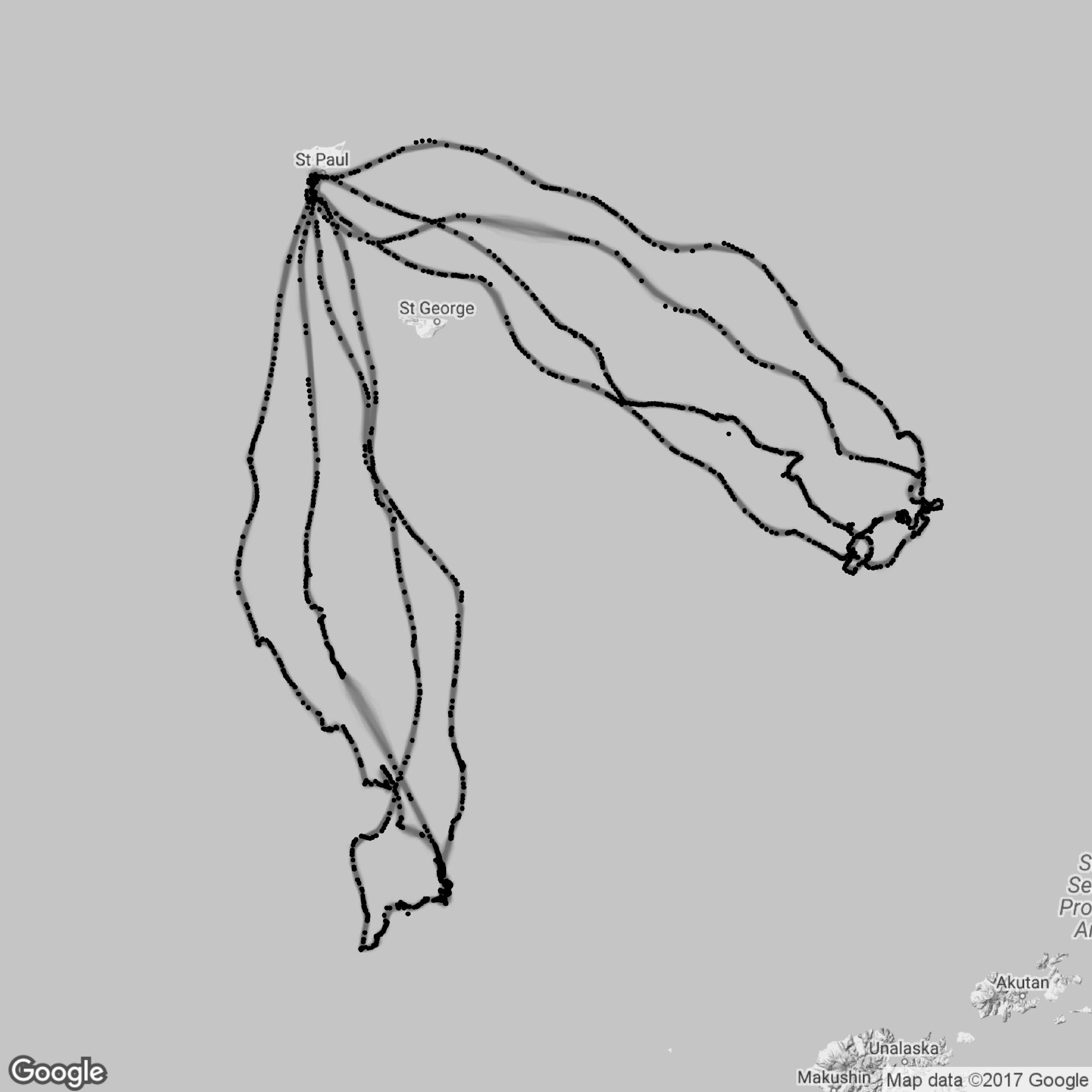}
  \caption{\footnotesize Movement of a female northern fur seal (\textit{Callorhinus ursinus}). The gray lines are draws from the approximate imputation distribution, $\lb \bmu^* | \bs \rb$. The black points are the data.}
  \label{fig:paths_nfs}
\end{figure}

Figure \ref{fig:paths_nfs} shows the observed locations of a northern fur seal (\textit{Callorhinus ursinus}) as it makes several foraging excursions from the Pribilof islands in Alaska. By imputing several samples from an approximation to the distribution of the true, continuous-time movement process, it is possible to obtain approximate inference from a movement model that would otherwise be too computationally expensive to fit (Section \ref{sec:application}).

Imputation strategies have been used in statistics for decades to address missing data issues. For example, a simple, yet \emph{ad hoc}, imputation procedure for telemetry data involves reconstructing the trajectory using linear interpolation (i.e., connecting the dots). The imputed trajectory is then treated as data in a statistical model that requires knowledge of the true unknown continuous path. This procedure is usually inadequate for several reasons: (1) the linearly interpolated path is not based on an understanding of the dynamics in the trajectory and, thus, provides a biased understanding of the individual's position between observed time points, (2) the telemetry data are considered to be perfectly accurate observations of the trajectory at the observation times, and (3) the uncertainty in our understanding of the trajectory between (or at) observation times is not accommodated.

Multiple imputation for random trajectories was developed to help alleviate some of these issues \citep{Rubin1996, Hooten2010, Hanks2011}. By imputing several, model-based paths, it is possible to account for underlying movement dynamics and measurement error. In what follows, we review imputation strategies from a Bayesian perspective. We then show how to use imputation for fitting continuous-time continuous-animal space movement models \citep[see][for the discrete-time setting]{Scharf2016, McClintock2017}. We examine the strengths and weaknesses of imputation approaches empirically, using simulation. Finally, we demonstrate imputation while modeling real telemetry data for a northern fur seal.

\section{Methods}
\subsection{Multiple imputation}
Conventionally, statistical methods for imputation were developed to address issues with missing data. Prior to more formal multiple imputation approaches \citep[e.g.,][]{Rubin1996}, a suite of \emph{ad hoc} methods were proposed (e.g., complete-case analysis, available-case analysis, fill-in with means; \citealt{Little1987}). In the Bayesian context, the basic idea behind imputation is that inference for the model parameters $\btheta$ under the observed data can be obtained using the posterior
\begin{align}
  \lb \btheta | \bs \rb  &= \int \lb \btheta , \bs_m | \bs \rb  d\bs_m, \\
  &= \int \lb \btheta | \bs, \bs_m \rb \lb \bs_m | \bs \rb  d\bs_m, \label{eqn:dataimp}
\end{align}
where, $\bs_m$ represents the missing data, and, following \cite{Gelfand1990}, we use brackets to denote a probability distribution. The distribution $\lb \btheta | \bs, \bs_m\rb $ is often referred to as the ``complete-data'' posterior and the distribution $\lb \bs_m | \bs \rb $ is the distribution of the missing data, conditioned only on the observed data (i.e., the imputation distribution). Thus, heuristically, the desired posterior distribution $\lb \btheta | \bs\rb $ is the complete-data posterior distribution averaged over the imputation distribution \citep{Rubin2004}.

The Bayesian imputation concept yields useful properties of posterior moments. In particular, the posterior mean and variance of the parameters can be written using the iterated expectation and variance formulations:
\begin{align}
  \E \lp \btheta | \bs \rp = \E \lp \E \lp \btheta | \bs, \bs_m \rp |\bs \rp,
\end{align}
and
\begin{align}
  \Var \lp \btheta | \bs \rp = \E \lp \Var \lp \btheta | \bs, \bs_m \rp | \bs \rp +
  \Var \lp \E \lp \btheta | \bs, \bs_m \rp | \bs \rp.
\end{align}
In practice, for a given observed data set, if $K$ samples of the missing data can be obtained from the imputation distribution, then the posterior expectation can be approximated by fitting the complete-data models to each complete-data set and then averaging the posterior means. Similarly, by averaging the posterior variances from the complete-data models and adding the sample variance of the posterior means, we arrive at an approximation of the correct posterior variance $\Var \lp \btheta | \bs \rp$. These approximations improve as the number of missing data samples from the imputation distribution ($K$) increases.

Multiple imputation depends on the ability to evaluate the complete-data posterior, and the ability to sample missing data sets from the imputation distribution. In the context of animal movement, telemetry data are measurements of the true positions with error and models for movement dynamics are most commonly specified for the true position process. By treating the true position process as missing data and taking a Bayesian multiple imputation approach, it is possible to obtain approximate inference under models that we would otherwise be unable to use.

\subsection{Animal movement}\label{sec:animal_movement}
To introduce the concept of imputation in a continuous-time animal movement setting, we denote the set of telemetry data as $\bs\equiv (\bs(t_1)',\ldots,\bs(t_n)')'$, a $2n \times 1$ vector, for observation times $t_1, \ldots, t_n$. In this setting, each $\bs(t_i)$ is a $2\times 1$ vector representing an observed telemetry position in two-dimensional space at time $t_i$. Similarly, we represent the true latent position process as $\bmu \equiv \lp \bmu(t_1)',\ldots,\bmu(t_m)' \rp'$, a $2m \times 1$ vector where each $\bmu(t_j)$ represents a true, but unknown, position at time $t_j$.

The true position process is typically modeled as arising from a distribution conditioned on a set of process model parameters $\btheta$, such that $\bmu \sim \lb \bmu | \btheta\rb $. In a hierarchical framework, the telemetry data are often modeled as conditional on the true latent trajectory and observation parameters, $\bpsi$, as $\bs \sim \lb \bs | \bmu, \bpsi \rb $. The hierarchical framework is useful for analyzing telemetry data when it is reasonable to assume that the processes that gives rise to measurement error is conditionally independent from the processes that describe the behavior of the animal.

An important class of models for animal movement that fits into our hierarchical framework can be described using sets of stochastic differential equations (SDEs). \cite{Brillinger1998} introduced one such model to study the migration of northern elephant seals that explicitly accounted for the curvature of the Earth. A general model that describes the motion of a Brownian particle in an external, conservative force field is given by the SDEs
\begin{align}
  \begin{split}
    d\bmu(t) &= \bv(t) dt \\
    d\bv(t) &= -\nabla H \lp \bmu(t), t \rp dt - \sigma_v\bv(t)dt + \sigma_v d\bb(t),
  \end{split} \label{eqn:SDE}
\end{align}
where $d\bb(t) \sim \N \lp \bzero, \; \bI_2 dt \rp$ and $H(\cdot, \cdot)$ is a time-varying potential function, the negative gradient of which yields the force acting on the Brownian particle \citep{Nelson1967}. In practice, a dense grid of times $t_1, \dots, t_m$ are used to fit the model, and each location, $\bmu(t_j)$, and velocity, $\bv(t_j)$, have an associated $dt_j = t_j - t_{j-1}$.

Modeling animal movement using \eqref{eqn:SDE} allows researchers to study the impacts of a wide variety of natural phenomena on an individual's behavior by specifying an appropriate potential function. For instance, if we define $H$ as
\begin{align}
  H(\bmu(t), t) = \beta(t) \|\bmu(t) - \bc \|_2, \label{eqn:potential_function}
\end{align}
where $\|\bmu(t) - \bc \|_2$ is the Euclidean distance between $\bmu(t)$ and $\bc$, and $\bc$ is a feature of interest in the environment, then the force on the particle is a pull toward $\bc$ for $\beta(t) > 0$, and repulsion from $\bc$ for $\beta(t) < 0$. The model in \eqref{eqn:SDE} -- \eqref{eqn:potential_function} provides an excellent description of the behavior of many marine mammals, such as northern fur seals. In summer months, female northern fur seals spend several days at sea foraging before returning to a specific rookery to nurse their pups \citep{Johnson2013}. By defining $\bc$ to be the location of the rookery, we can accommodate the time-varying effect it has on a nursing female's movement (see Section \ref{sec:application}).

To form a complete hierarchical model, we specify independent, circular Gaussian measurement error, and conjugate priors for $\sigma_s^2$ and the process parameters in $\btheta$ as
\begin{align}
    \bs(t) | \bmu(t) &\sim \N \lp \bmu(t), \; \sigma_s^2 \bI_2 \rp, \\
    \sigma_s^2 &\sim \IG(a_s, b_s), \\
    \sigma_v^2 &\sim \IG(a_v, b_v), \\
    \bbeta &= \bW \balpha, \\
    \balpha &\sim \N \lp \bzero, \; \sigma^2_\alpha \bI \rp
  \label{eqn:sde_priors}
\end{align}
where $\bW$ is a matrix of basis functions used to model the dynamic process $\bbeta = \lp \beta(t_1), \dots, \beta(t_m) \rp'$ (e.g., splines, \citealt{Hefley2016}). Fitting the SDE-based model to telemetry data using an MCMC algorithm is computationally intensive due to the non-linear relationship between $\bmu$ and $\btheta \equiv \lp \bbeta', \sigma_v^2 \rp'$. Standard methods require an MCMC algorithm in which $\bmu(t)$ must be updated over a fine grid of times, and for many modern telemetry data sets, the effort required to fit the model can exceed the computational resources available on even the most advanced high performance computing platforms. However, conditioned on the true path, $\bmu$, inference for $\btheta$ is straightforward.

\subsection{Process Imputation}
\cite{Hooten2010} and \cite{Hanks2011} proposed an approximate-model fitting procedure that is similar in spirit to multiple imputation, which we term ``process imputation.'' The motivation for the process imputation approach arises from a decomposition of the posterior distribution of the process parameters. Using standard properties of conditional probability, we write the posterior distribution as
\begin{align}
  \lb \btheta | \bs\rb  &= \int \lb \btheta | \bmu, \bs \rb \lb \bmu | \bs \rb d\bmu. \label{eqn:procimp}
\end{align}
The form of the integral in \eqref{eqn:procimp} is similar to that used in conventional data imputation approaches \eqref{eqn:dataimp}; however, in this case, the process $\bmu$ takes the place of the missing data $\bs_m$.

As with conventional imputation approaches, if we can obtain the posterior distribution $\lb \btheta | \bmu, \bs \rb$ and samples from the process imputation distribution $\lb \bmu | \bs \rb $, then we can take the same approach of fitting the process model to each of the imputed process samples and approximating posterior moments for $\btheta$ using the relevant iterated expectation and variance equations.

For many movement models, including both the SDE-based model introduced in \eqref{eqn:SDE}, and the discrete-space models presented by \cite{Hooten2010}, \cite{Hanks2015}, and \cite{Hanks2016}, evaluating $\lb \btheta | \bmu, \bs \rb$ up to proportionality is straightforward. The process model parameters, $\btheta$, are often conditionally independent of data, $\bs$, and the observation parameters, $\bpsi$, such that the complete-data posterior simplifies to $\lb \btheta | \bmu \rb \propto \lb \bmu | \btheta \rb \lb \btheta \rb$. In practice, the challenge to implementing a process imputation approach is in sampling from the imputation distribution, $\lb \bmu | \bs \rb$. The difficulty arises for the same reasons we are unable to fit the full hierarchical model. Sampling from the imputation distribution, $\lb \bmu | \bs \rb \propto \int \lb \bs | \bmu, \bpsi \rb \lb \bmu \rb \lb \bpsi \rb d\bpsi$, requires that we have access to the marginal distribution $\lb \bmu \rb$, which, in turn, requires we evaluate the integral $\lb \bmu \rb = \int \lb \bmu | \btheta \rb \lb \btheta \rb d\btheta$. For the models we consider, this integral is intractable, either for reasons involving noninvertibility or computation.

In what follows, we assume that direct sampling from $\lb \bmu | \bs \rb$ is impossible. To circumvent this obstacle, we assume the existence of another conditional random variable $\bmu^*$. The distribution of $\bmu^*$ is sufficiently similar to $\bmu$ that draws from the distribution $\lb \bmu^* | \bs \rb$, which may or may not depend on $\bpsi$, can be used to approximately evaluate the integral in \eqref{eqn:procimp}. We term the distribution given by $\lb \bmu^* | \bs \rb$ the ``approximate imputation distribution'' (AID), and denote the resulting approximation to the true posterior by $\lb \btheta^* | \bs \rb$. When implementing process imputation, selection of the AID is performed by first specifying a model for $\lb \bmu^* | \bs, \bphi \rb$ parameterized by $\bphi$. Estimates for the parameters $\bphi$ are obtained in a preliminary stage by fitting $\lb \bmu^* | \bs, \bphi \rb$ to the data. Common models for the AID used in the literature include the velocity-based Ornstein-Uhlenbeck model of \cite{Johnson2008} and the spline-based functional model of \cite{Buderman2016}.

As motivation for substituting draws for the true imputation distribution with draws from an approximate distribution, we offer the following heuristic argument based on asymptotically dense observations with negligible measurement error. Additionally, we discuss the results of two simulation studies in Section \ref{sec:simulation_study} that provide practical guidance for using a process imputation approach.

When observations with small measurement error are available, the distribution $\lb \bmu | \bs \rb$ has probability mass concentrated on paths passing through the immediate vicinity of the telemetry locations, $\bs$. For the extreme case in which an observation $\bs(t)$ exists for all possible times $t$, and measurements are made without error, we have $\bs(t) = \bmu(t)$ for all $t$, and the conditional distribution $\lb \bmu | \bs \rb$ collapses to a point mass on the true path. At this limit, the distribution $\lb \bmu \rb$ could, in principle, be replaced by any probability measure with the correct support without affecting $\lb \bmu | \bs \rb$. Thus, in the limiting case of infinitely dense observation times and zero measurement error, any AID that concentrates probability mass around paths through the data is sufficient. Following this argument, it is reasonable to expect that the degree to which substituting $\lb \bmu^* | \bs \rb$ in equation \eqref{eqn:procimp} is an adequate approximation will depend on the density of observation times and the severity of measurement error.

An analogy to variational approximations \citep{Ormerod2010} provides additional motivation for the process imputation approach. Variational approximations are often employed in a Bayesian setting to find approximations to posterior densities when sampling from the true densities is computationally infeasible. Optimal approximations are derived by minimizing the Kullback-Leibler divergence between a family of proposal distributions (e.g., distributions for which certain model parameters are assumed to be independent) and the true distribution. In the process imputation setting, the model for the AID takes on the role of the family of proposal distributions. Instead of minimizing the Kullback-Leibler divergence, we select the optimal approximation by fitting the AID model to the data. The accuracy of a variational approximation to the true posterior density depends, in part, upon the suitability of the family of proposal distributions. Our simulation study suggests that the choice of AID model in the process imputation setting has an analogous relationship with the approximate posterior distribution, $\lb \btheta^* | \bs \rb$.

In practice, when measurement errors are non-zero and the number of observations is finite, the model for the AID can have a significant impact on inference. When implementing this procedure, a finite number of draws from the AID are used to compute the integral in \eqref{eqn:procimp}; using too few draws can fail to adequately represent the AID, biasing inference. We investigated the effects of each of these factors on inference obtained for the SDE-based movement model using process imputation with a simulation study (Section \ref{sec:simulation_study}).

\section{Simulation studies}\label{sec:simulation_study}
We conducted two simulation studies of the process imputation approach in the context of SDE-based models for movement. The results revealed important information about when the approach may be useful, and offer insight into the viability of process imputation for other movement models.

\subsection{Study 1: Brownian particle in an external force field}\label{sec:sim1}

\begin{figure}[H]
\centering
\includegraphics[width = 0.8\linewidth]{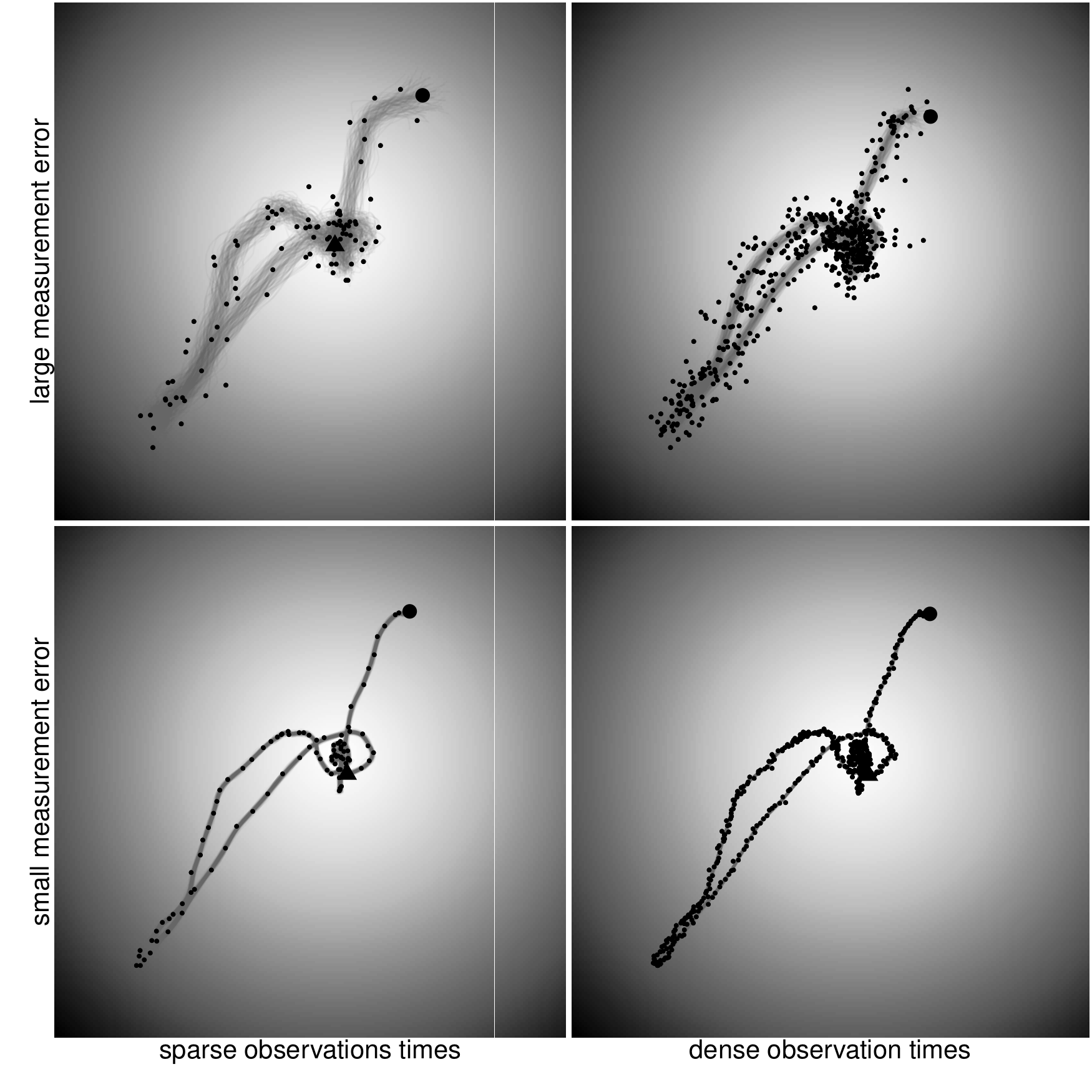}
\caption{\footnotesize Examples of simulated data from the four different combinations of measurement error severity and observation frequency. Points represent the simulated data, gray lines represent samples from an AID, and the large circle and triangle represent the first and last observations, respectively. The background represents the potential surface, $H$. For positive values of $\beta(t)$, darker shades represent low values, and lighter shades represent high values.}
\label{fig:me_ot_eg}
\end{figure}

The first simulation study was designed to explore the impacts of four different factors on the validity of inference obtained for the movement model specified in \eqref{eqn:SDE} -- \eqref{eqn:sde_priors} using process imputation: 1) the severity of measurement error, 2) the temporal density of observations, 3) the model for the AID, and 4) the number of draws, $K$, from the AID used in the analysis. Figure \ref{fig:me_ot_eg} shows examples of simulated data (black points) along with imputed paths (gray curves) for the different combinations of measurement error severity and observation frequency. For all four combinations of measurement error (``large'' and ``small'') and observation frequency (``sparse'' and ``dense''), we simulated 24 paths from the SDE-based movement model (see Appendix A for details). We then fit two different AID models to the simulated data: the integrated Ornstein-Uhlenbeck model of \cite{Johnson2008} using the \verb=R= package \verb=crawl= (``OU''), and a Gaussian process \citep[e.g.,][]{Fleming2015, Hooten2016} with Gaussian covariance function (``GP'') (see Appendix A for further details). Finally, for each combination of measurement error, observation frequency, and AID model, we drew $K$ realizations from the AID to approximate the integral \eqref{eqn:procimp}. We investigated three different values for $K$ (8, 32, and 128), and also implemented the process imputation approach using the posterior mean of the AID. To provide a baseline against which to compare the performance of process imputation, we also fit the SDE-based model conditioned on the true, underlying path $\bmu$.

\subsubsection{Implementation}
After a suitable AID has been identified, the process imputation approach can be implemented using an MCMC algorithm. We obtained samples from the approximation to the true posterior distribution, $\lb \btheta^* | \bs \rb$, by drawing realizations form the complete-data posterior, $\lb \btheta | \bmu \rb$, and integrating with the respect to the AID. The two steps can be combined into a single MCMC algorithm using the following procedure:
\begin{enumerate}
\item Draw realizations from the AID, ${\bmu^*}^{(k)} \sim \lb \bmu^* | \bs \rb$, for $k = 1, \dots, K$.
\item MCMC procedure:
  \begin{enumerate}
  \item Randomly select one of the $K$ samples, ${\bmu^*}^{(k)}$, with probability $1/K$. \label{itm:select_k}
  \item Update model parameters $\btheta | \bmu = {\bmu^*}^{(k)}$ conditioned on the imputed path, ${\bmu^*}^{(k)}$. \label{itm:update}
  \item Repeat steps 2.(a) - 2.(b) at each iteration of the MCMC algorithm.
  \end{enumerate}
\end{enumerate} \label{enumerate_steps}

\subsubsection{Results}
\begin{figure}[H]
  \centering
  \includegraphics[width = 0.8\textwidth]{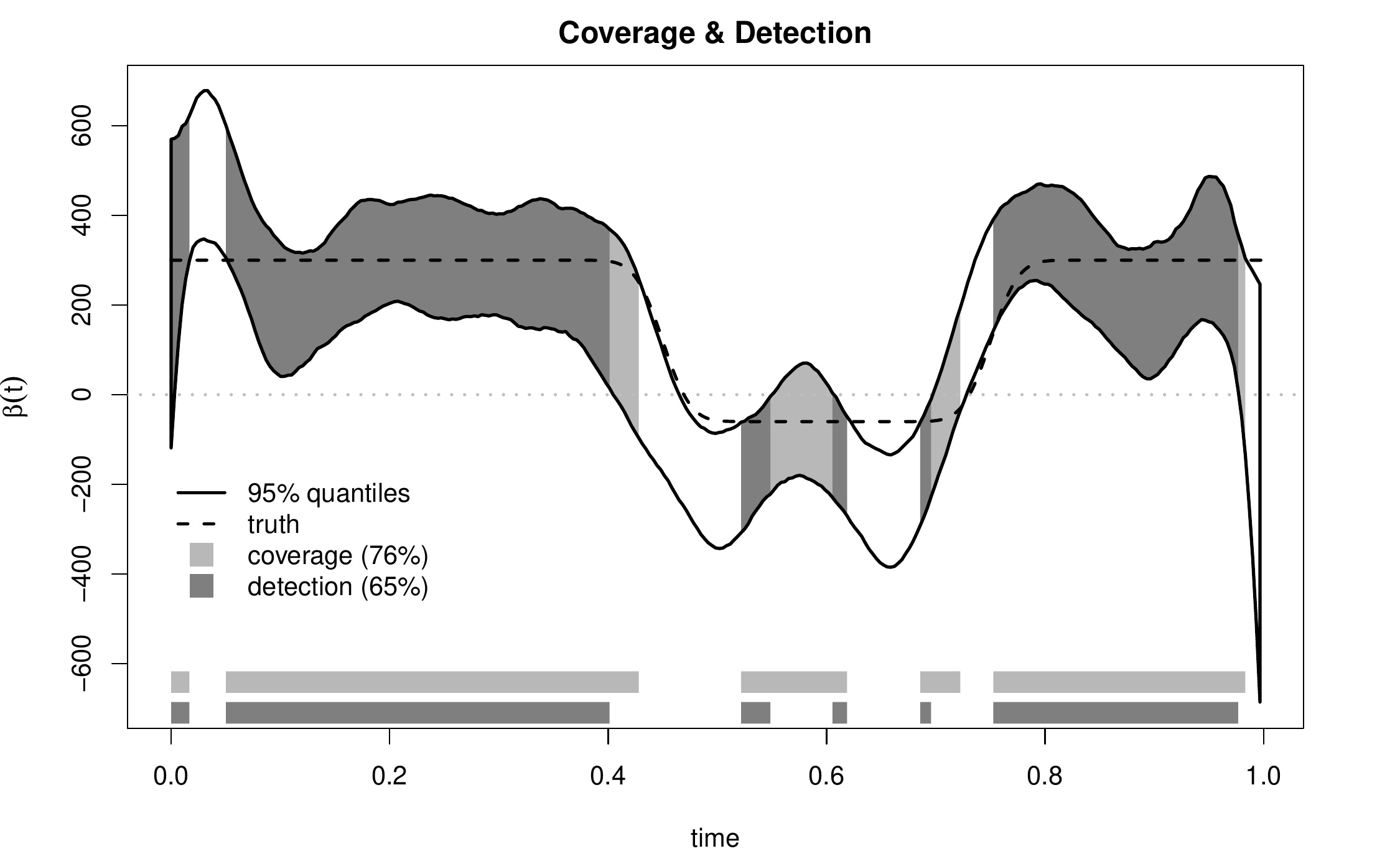}
  \caption{\footnotesize Proportion of $\bbeta$ covered and detected by the equal-tailed 95\% credible interval. The ``coverage'' region (light gray) is defined by times at which the truth falls within the 95\% credible interval (black lines). The ``detection'' region (dark gray) is a subset of the coverage region, and occurs when the 95\% credible does not contain zero.}
  \label{fig:cd_eg}
\end{figure}

To evaluate the performance of the process imputation approach at estimating the effect of the potential function, we compared the posterior distributions of $\bbeta$ to the values used to simulate the data through two summary measures. We assessed the accuracy of process imputation by computing the proportion of the study period for which the pointwise, equal-tailed, 95\% credible interval contains the true value of $\beta(t)$, which we term the ``coverage'' region. We assessed the precision of process imputation by computing the proportion of the study period during which the pointwise credible interval both contains the truth, and does not contain 0, which we term the ``detection'' region. Thus, evidence of strong performance by the process imputation approach is characterized by a combination of large coverage and detection regions. Figure \ref{fig:cd_eg} shows an example of coverage and detection regions. For further details concerning coverage and detection regions, see Appendix A. To evaluate the performance of the process imputation approach at estimating $\sigma_v^2$ and $\sigma_s^2$, we computed the proportion of simulations in which equal-tailed 95\% credible intervals contained the true values of the parameters. 

\begin{figure}[H]
  \centering
  \subfigure{\includegraphics[width = 0.48\linewidth]{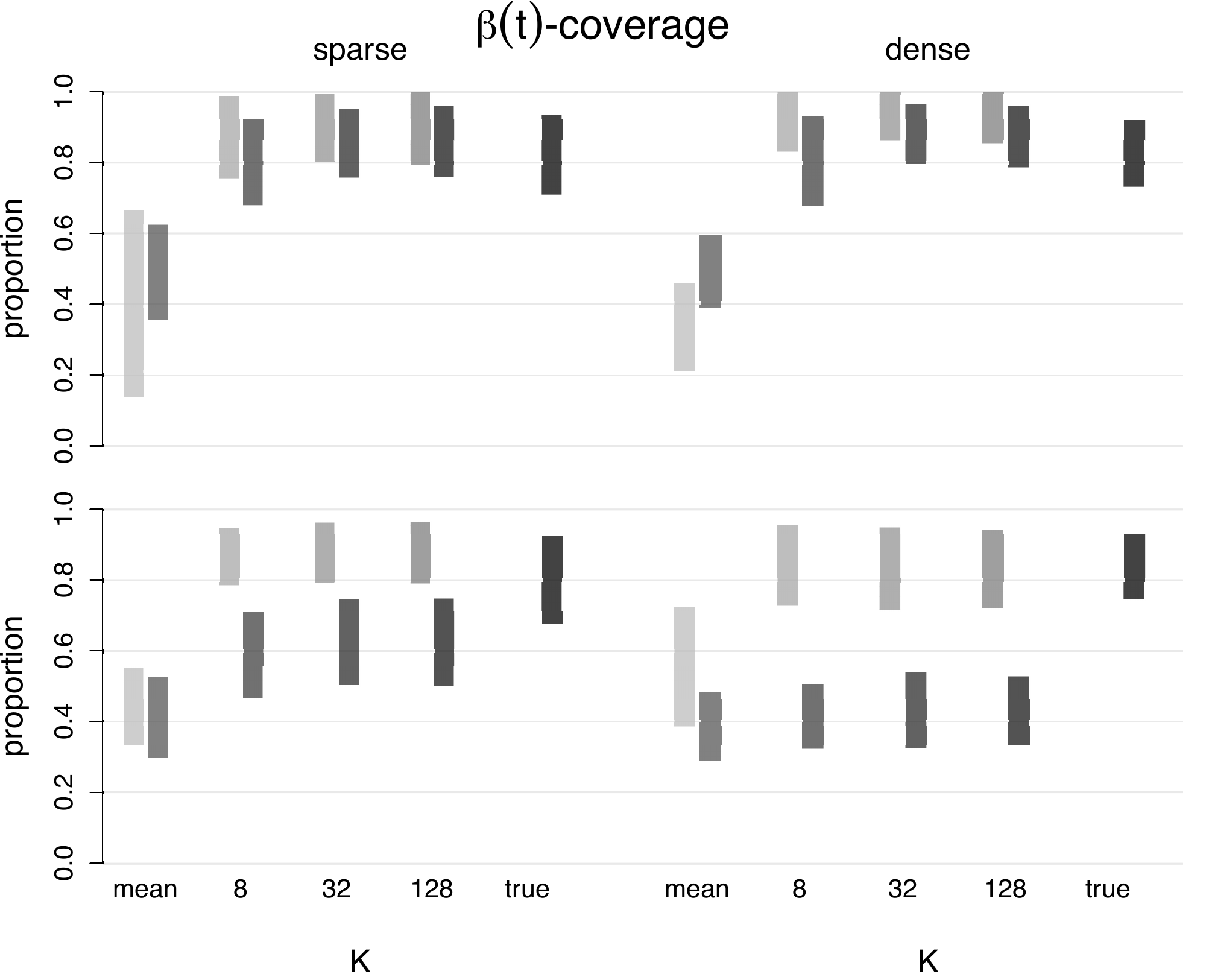}}
  \subfigure{\includegraphics[width = 0.48\linewidth]{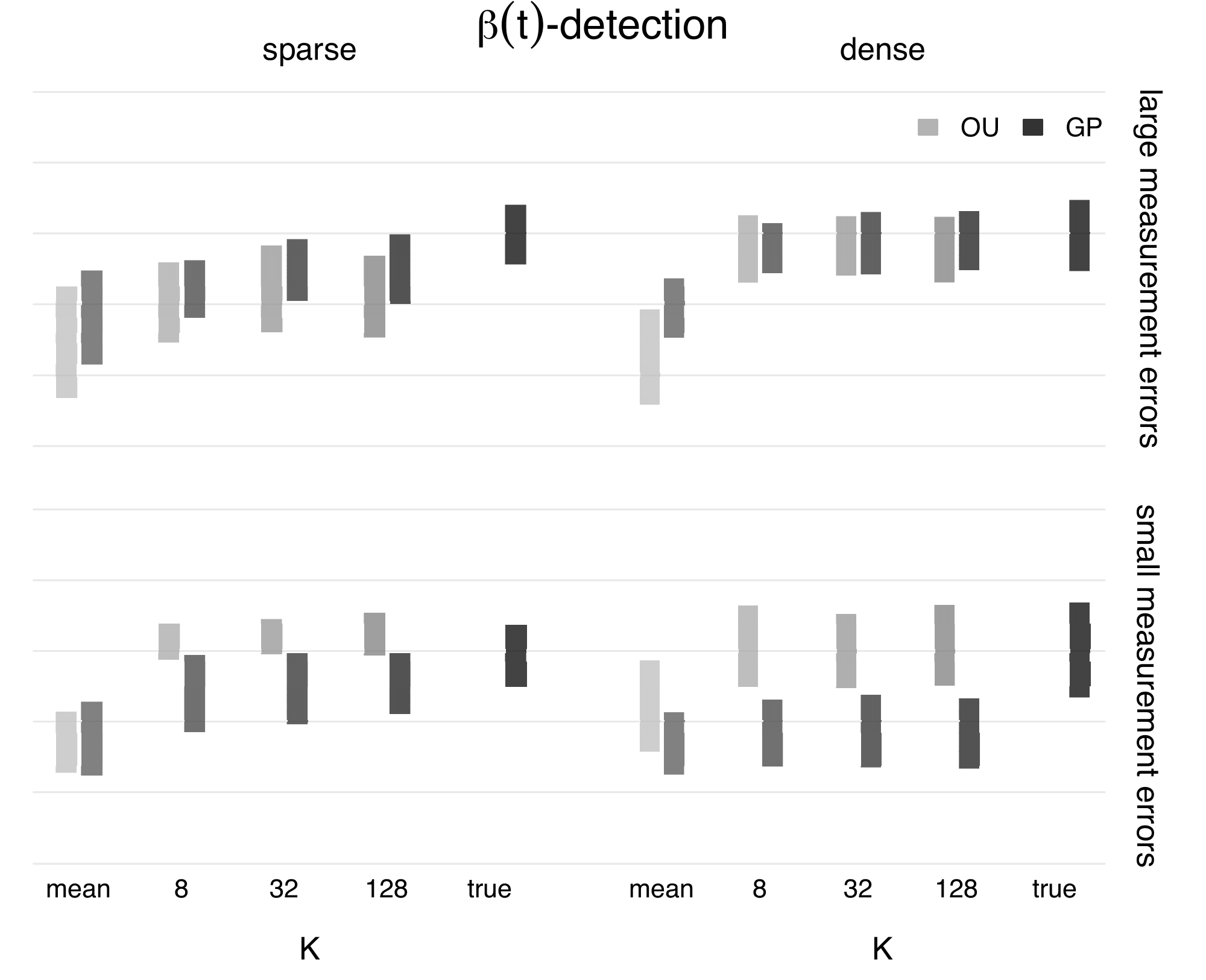}}
  \caption{\footnotesize Proportion of $\bbeta$ covered/detected by the equal-tailed 95\% credible interval. The range of each bar represents the 12.5\% and 87.5\% quantiles across the 24 simulations. See Figure \ref{fig:cd_eg} for details on the definition of the ``coverage'' region.}
  \label{fig:sim_beta_cover_detect}
\end{figure}

\begin{figure}[H]
  \centering
  \includegraphics[width = 0.5\textwidth]{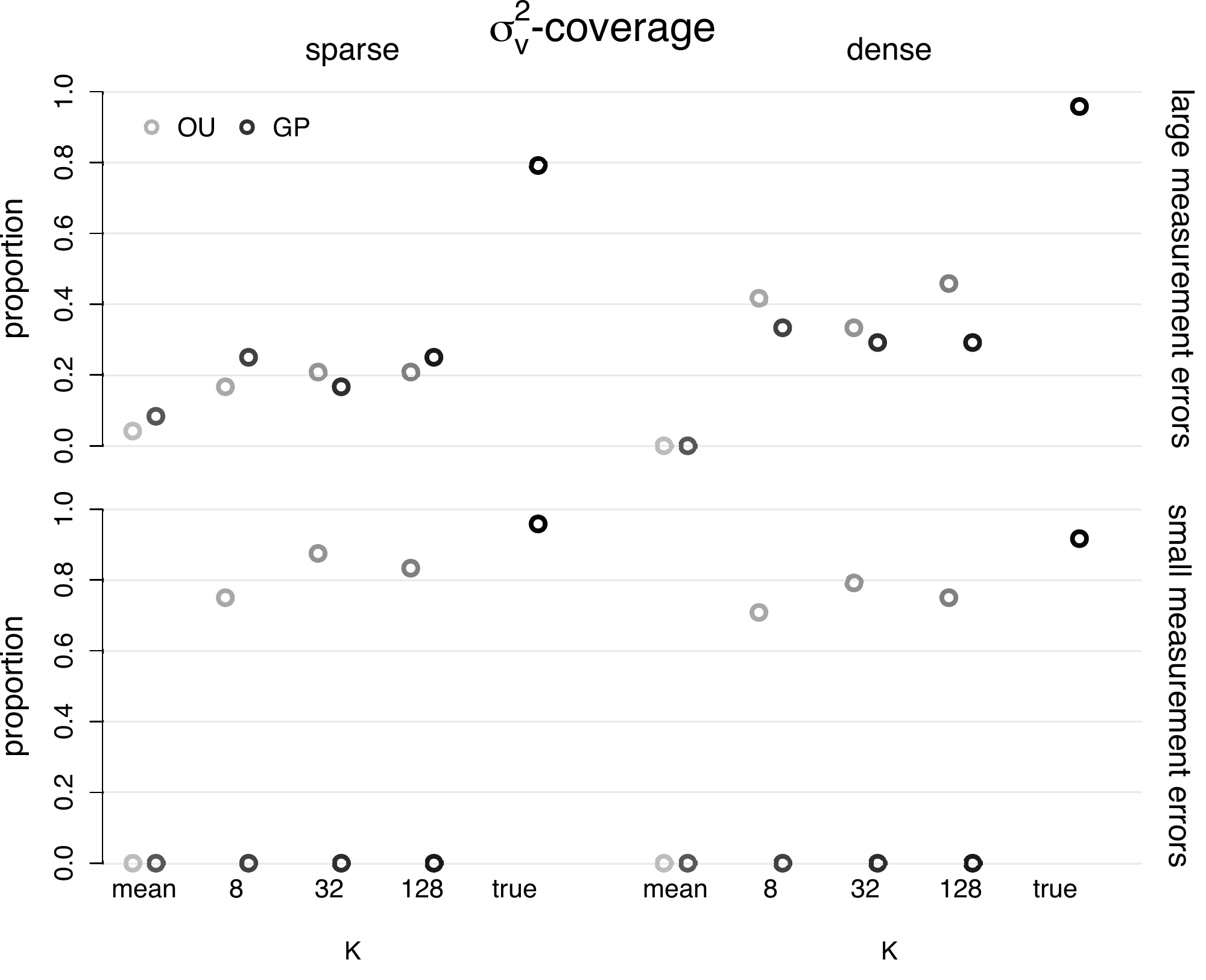}
  \caption{\footnotesize Proportion of 24 paths with 95\% credible intervals that cover the true value of $\sigma_v^2$.}
  \label{fig:sim_sigsqv_cover}
\end{figure}

Figure \ref{fig:sim_beta_cover_detect} shows the sizes of coverage and detection regions under each process imputation regime, repectively. The vertical range of the bars shows the 12.5\% to 87.5\% quantiles across each set of 24 simulations. Lighter shades correspond to cases for which the AID model is the Ornstein-Uhlenbeck model of \cite{Johnson2008}, and darker shades correspond to cases for which the AID model is a Gaussian process. Rows and columns organize cases by measurement error and observation frequency, respectively. Finally, within each plot, cases are organized by the number of draws used from the imputation distribution in fitting the approximate hierarchical model. Figure \ref{fig:sim_sigsqv_cover} shows the proportion of the 24 simulated paths for which the equal-tailed 95\% credible interval contained the true value of $\sigma_v^2$, organized analogously. The far left of each subplot in Figures \ref{fig:sim_beta_cover_detect} and \ref{fig:sim_sigsqv_cover} corresponds to the case of a single imputed path that is the posterior mean of the AID, and the far right corresponds to the case of a single imputation of the true, simulated path. Thus, the far right values in each subplot represent a baseline against which we can compare the performance of the other regimes. Figure 8 in Appendix A shows the coverage proportions for $\sigma_s^2$.



From Figure \ref{fig:sim_beta_cover_detect}, it is clear that merely imputing the posterior mean of the AID, rather than several samples, does not adequately represent the variability of the imputation distribution, leading to bias and poor coverage of the true function $\bbeta$. However, using a small number of imputed paths is generally sufficient to achieve a large coverage region.

Second, in all cases, imputing from the OU-based AID yields larger coverage regions as compared to the GP-based AID. A large coverage region is not, by itself, evidence of strong performance. However, the OU-based AID also generally yields larger detection regions, suggesting that the OU-based AID better represents the true imputation distribution than the GP-based AID, at least for the particular case of the SDE-based movement model defined by \eqref{eqn:SDE} -- \eqref{eqn:sde_priors}.

Third, the sizes of the coverage and detection regions for cases corresponding to an OU AID model are generally insensitive to measurement error and observation frequency (Figure \ref{fig:sim_beta_cover_detect}), with the only exception being the case of high measurement error and sparse observations. The same cannot be said of the GP AID model, for which coverage and detection of $\bbeta$ deteriorates when measurement errors are small, especially when observations are dense in time. The weakness of the GP AID model for the case of small measurement error is apparent in Figure \ref{fig:sim_sigsqv_cover} where coverage drops to zero.

Finally, inference for the function $\bbeta$ using the process imputation approach appears to be more robust across different values of measurement error, observation frequency, and AID models than inference for $\sigma_v^2$. Neither AID model resulted in adequate coverage for $\sigma_v^2$, with the possible exception of the OU model for the cases corresponding to small measurement error. Therefore, the results of the study suggest that process imputation is a reasonable approach for obtaining inference about the effect of the potential surface, $H$, which depends directly on the position of the individual. However, there is also evidence that process imputation may not be a valid approach for obtaining inference about the parameter $\sigma_v^2$, which relates to the velocity of the individual.

\subsection{Study 2: Comparing exact and two-stage approaches}\label{sec:sim2}

The simulation study discussed in \ref{sec:sim1} is useful because it demonstrates the viability of the two-stage approach to approximate inference for the case when exact inference is computationally infeasible. By the same token, we cannot compare the approximate inference from the two-stage procedure to exact inference. Any model for which we could obtain both forms of inference must necessarily be somewhat contrived, because the very accessibility of exact inference negates the need for an approximate procedure. Still, it is instructive to consider a simplified setting when both exact and approximate inference is available to see what differences exist.

We investigate the model employed by \cite{Brillinger1998} for the continuous movement of a particle under the influence of an external force. As earlier, let $\bmu(t)$ denote the location of a particle in two dimensions at time $t$. Let $\bH(\bmu(t), \beta)$ be a potential surface whose negative gradient represents the force operating on the particle, the value of which may depend upon the parameter $\beta$. In contrast to the model used in the first simulation study, we assume the parameter $\beta$ is constant in time. The continuous path of the particle is modeled with the stochastic differential equation
\begin{align*}
d\bmu(t) &= -\nabla \bH(\bmu(t), \beta) dt + d\bb(t) \\ \label{eqn:brillinger}
\bmu(0)|\sigma_0^2 &\sim \N\lp \bzero, \sigma_0^2 \bI_2\rp,
\end{align*}
where $d\bb(t)$ are infinitesimal increments of Brownian motion. As before, we model the measured locations, $\bs(t)$, of the particle as conditionally Gaussian with mean equal to the true location, $\bmu(t)$, and variance $\sigma_s^2$.

An important distinction between this model and that presented in Section \ref{sec:simulation_study} is that the ``force'' defined by the potential function now operates directly on the position process, rather than the velocity. Therefore, interpretation of the effect in the context of classical Newtonian physics is no longer possible, and continuing to refer to the function $\bH$ as a potential function and its gradient as a force may be misleading. Additionally, paths generated from this model do not have well-defined velocities at all time points, and may not be realistic models for the movement of massive bodies.

Nevertheless, the stochastic differential equation is sometimes sufficient for approximating the movement of particles under the influence of an external force. We consider the case when $\bH(\bmu(t), \beta) = \beta \| \bmu(t) - \bc\|_2$, as in Section \ref{sec:animal_movement}, such that the particle feels a constant ``push'' toward the location $\bc$. 

Similar to the simulation study in Section \ref{sec:sim1}, we fit the model to simulated data using the proposed two-stage approximation. For the purposes of this simulation study, we used the Ornstein-Uhlenbeck model of \cite{Johnson2008} for the AID and obtained inference using the posterior mean of the fitted AID, as well as $K=8, 32,$ and $128$ draws from the AID. Additionally, we obtained exact inference for the model in an MCMC framework by incrementally updating the true, unobserved path $\bmu$ over a dense grid of time points $t_1, \dots, t_m$ (see Appendix A for further details). Settings with both ``sparse'' and ``dense'' rates of observation (100, 500 observation times, respectively), as well as ``large'' and ``small'' measurement errors ($\sigma_s^2 = 10^{-2}, 10^{-4}$ respectively) were considered (see Figure 9 in Appendix A for examples). 

\begin{figure}[H]
  \centering
  \subfigure{\includegraphics[width = 0.48\textwidth]{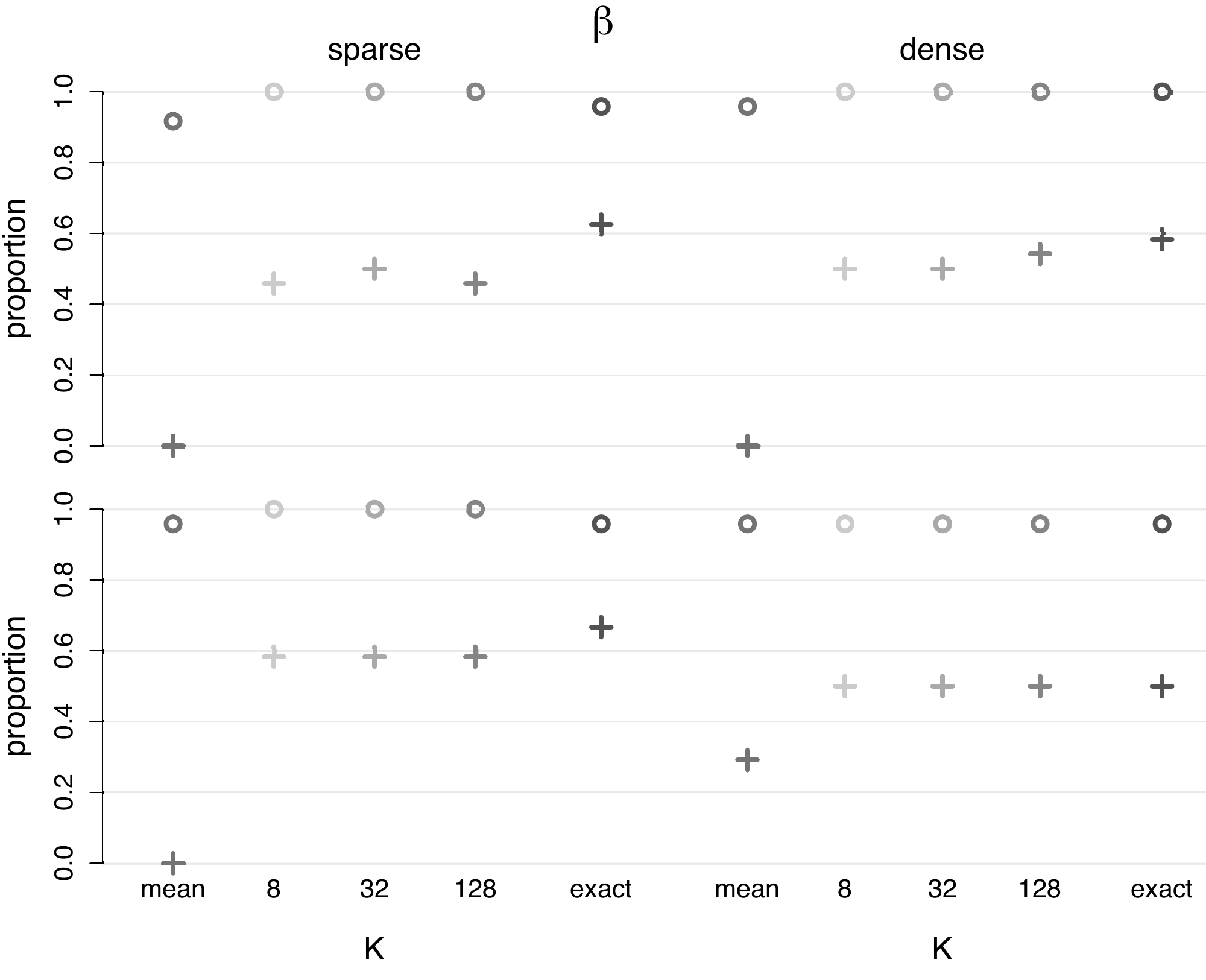}}
  \subfigure{\includegraphics[width = 0.48\textwidth]{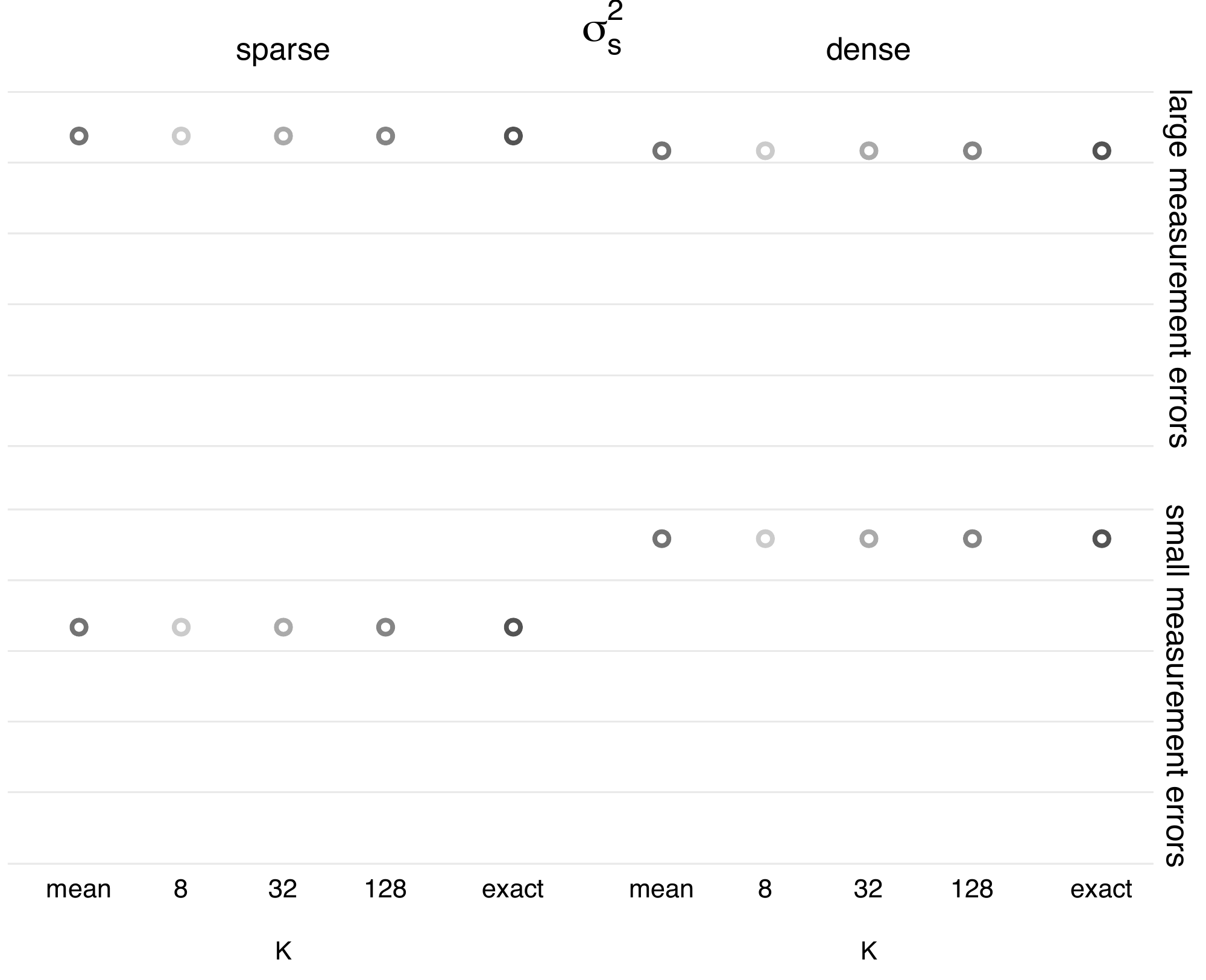}}
  \caption{\footnotesize Proportion of 24 simulated paths for which 95\% credible intervals covered/detected the true value. The four left plots correspond to coverage/detection of $\beta$. The ``o'' symbols represent coverage, and the ``+'' symbols represent detection. The four right plots correspond to coverage of $\sigma_s^2$.}
  \label{fig:cd_brill_beta_sigsq_s}
\end{figure}

We evaluated the performance of each method by examining coverage and detection rates (see Section \ref{sec:simulation_study}) across 24 different simulated paths. The left half of Figure \ref{fig:cd_brill_beta_sigsq_s} shows proportion of simulated paths for which the equal-tailed 95\% posterior credible interval overlapped the true value (``o''), and the proportion of simulated paths for which the credible intervals both overlapped the true value, and failed to contain 0 (``+''). Coverage was similar for both methods, across all choices of $K$, however detection was low for the process imputation approach when only the posterior mean of the AID was used. When 8 or more draws from the AID were used, detection rates for the two-stage process imputation method was similar to those obtained from exact inference, although for the case of ``sparse'' observation times, detection rates for the process imputation approach were slightly lower than those obtained for exact inference, suggesting that the two-stage procedure may result in inflated uncertainty about $\beta$ when the density of observations is low.

The right half of Figure \ref{fig:cd_brill_beta_sigsq_s} shows the coverage rates for the measurement error variance, $\sigma_s^2$. All methods performed similarly. The findings of the simulation study suggest that the two-stage approach is a viable option when multiple draws from the AID are used.

\section{Application}\label{sec:application}

We analyzed telemetry data gathered for a single female northern fur seal during the fall of 2008 (Figure \ref{fig:paths_nfs}). Observations were collected using high precision GPS devices at intervals of approximately 15 minutes (for more details see \cite{Johnson2013}) over the course of 6 weeks. The individual under study makes several long foraging excursions, returning approximately every 7 days to nurse pups at a rookery on Saint Paul Island. A vignette giving a detailed description of the model fitting procedure is provided in Appendix C.

We used process imputation to fit the SDE-based model given by \eqref{eqn:SDE} -- \eqref{eqn:sde_priors}, defining the center of attraction, $\bc$, to be the location of the rookery, $(57.13^\circ \text{N}, 170.27^\circ \text{W})$. Informed by the results of the simulation study (Section \ref{sec:simulation_study}), we selected the Ornstein-Uhlenbeck process of \cite{Johnson2008} as the AID model, and used $K=128$ imputed paths.

\begin{figure}[ht]
  \centering
  \includegraphics[width = 0.8\textwidth]{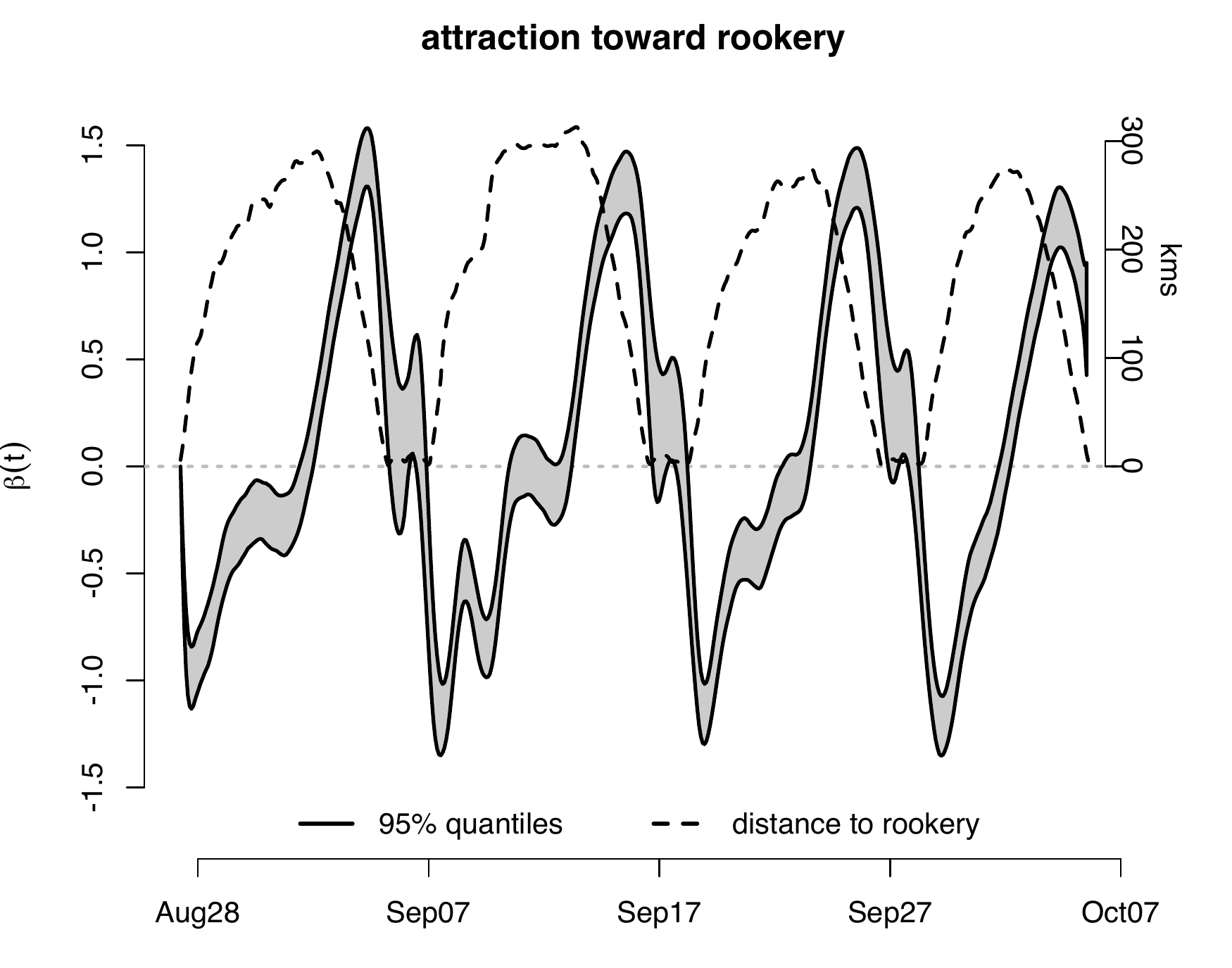}
  \caption{\footnotesize Time varying effect of distance to rookery. The gray polygon represents the pointwise equal-tailed 95\% credible interval for $\bbeta$, and the dashed line shows the distance from the animal's position to the rookery.}
  \label{fig:beta_nfs}
\end{figure}

Figure \ref{fig:beta_nfs} shows a pointwise equal-tailed 95\% credible interval for the dynamic effect of the rookery on the movement of the fur seal. The dashed line shows the distance the animal was from the rookery over time. The periodic pattern of repulsion ($\beta(t) < 0$) and attraction ($\beta(t) > 0$) that the fur seal experiences with respect to the rookery coincide with foraging excursions. A subtler pattern of behavior is visible in the fine scale oscillations that occur during the individual's travel away from the rookery (e.g., from approximately September 7 -- 10). During this phase of the foraging excursions, the force of repulsion away from the rookery is varying, indicating movement that is less directed and more exploratory. In contrast, during the second half of each foraging excursion (e.g., approximately September 12 -- 16), the force of attraction steadily increases until just before arrival, indicating strongly directed movement.

Several other relevant questions of scientific interest related to the patterns observed in $\bbeta$ can be addressed using approximate inference from the SDE-based model. Examples of how this model could be employed include: (1) the study of patterns in $\bbeta$ over the course of a long time period to learn about the physical development of northern fur seal pups as demonstrated by their mother's changing behavior; (2) investigation of models with additional potential surface terms for other factors from which researchers could learn about the effects of important environmental covariates on movement, while controlling for the effect of the rookery; and (3) comparison of the function $\bbeta$ across multiple animals to learn about individual-level variability among nursing females. Without using process imputation, mechanistically-motivated models such as this one are inaccessible to researchers studying animal movement.

\section{Discussion}

A key result from our simulation study for the case for SDE-based movement models is that process imputation is an effective approach for studying some types of model parameters, but not necessarily others. The approximate posterior distribution for the function $\bbeta$ is reasonably robust to the severity of measurement error, observation frequency, and AID model, as evidenced by the close agreement in coverage and detection between inference using process imputation and inference conditioned on the true movement process $\bmu$ (Figure \ref{fig:sim_beta_cover_detect}). In contrast, approximate inference for the parameter $\sigma_v^2$ varies substantially with the choice of AID model, and appears to be strongly dependent on the severity of measurement error and observation frequency (Figure \ref{fig:sim_sigsqv_cover}).

A practical interest for researchers is understanding the types of variables for which approximate inference using process imputation will be reliable. We suggest one possible indicator for reliability based on a distinction between ``position-order'' and ``higher-order'' effects on movement.

Generally, we define position-order effects on movement to be those directly related to the position process of an individual. We define higher-order effects to be those related to derivative processes such as velocity or acceleration. For the SDE-based movement model, the magnitude and direction of the force induced by the potential surface is defined by the the particle's location in space and is therefore primarily a position-order effect. Alternatively, the influence of the parameter $\sigma_v^2$ is through the particle's velocity rather than its location, and is therefore a higher-order effect. Telemetry data typically contain large amounts of information about an individual's position. However, information in the data about a particle's higher-order processes can be far less precise and deteriorates much more rapidly with increasing measurement error and gaps between observations. Therefore, we can expect parameters such as $\sigma_v^2$ to be more difficult to estimate and more strongly influenced by the choice of AID model in a process imputation framework. Currently, process imputation-based inference for higher-order parameters may be unreliable. However, many modern telemetry devices now collect direct measurements of an individual's velocity and acceleration, which may allow for a broader use of process imputation in the future. Additionally, the primary parameter of scientific interest in the SDE-based movement model is $\bbeta$; bias in the estimation of higher-order parameters such as $\sigma_v^2$ may be ignorable when it does not affect ecological learning.

In selecting an AID, the goal is to represent the distribution of the true path conditioned on the data as faithfully as possible. Thus, a first step is to ensure that candidate AIDs generate paths that are physically proximate to those arising from the true imputation distribution. Researchers should verify that the paths from candidate AIDs lie near the observed data. A greater challenge is to verify that the form of temporal dependence present in the AID is similar to that of the true imputation distribution, because the structure of dependence can vary considerably without affecting the central measure. For example, the OU and GP AIDs implemented in the simulation study in Section \ref{sec:simulation_study} generate paths with very different levels of smoothness, in the sense that the former has no continuous derivatives, while the latter has an infinite number. As a general rule, one should select a model for the AID whose dependence structure is similar to that of the process model. In practice, the family of models available for the AID will be limited by various constraints, including computational ones, and the researcher will be required to balance several competing priorities. For many applications, it may be helpful to consider multiple potential AIDs and investigate the sensitivity of all parameters of interest.

A common family of movement models that make use of process imputation are discrete-space models \citep{Hooten2010, Hanks2015, Hooten2016a, Hanks2016}, which model an individual's movement as a series of transitions between areal regions. Each transition is modeled using a multinomial distribution, where the probability of an individual moving to an adjacent areal region is a function of the associated environmental covariates. The effects of the environmental covariates on these transition probabilities are examples of position-order effects because they depend explicitly on the location of the individual (i.e., the areal region currently occupied). While the transition probabilities also depend on the residence time of an individual within each grid cell, and thus the average velocity of the animal during its residence, they do not depend explicitly on the instantaneous velocity at each time point. Our simulation study did not directly investigate the performance of process imputation for discrete-space movement models, but our findings suggest that process imputation may provide reliable inference about the relationship between environmental covariates and the inter-cell transition probabilities.

Although imputation-based approaches are an increasingly common tool for fitting models of animal movement to data, we are not aware of any other studies of their impact on inference. Our findings provide useful insight about the circumstances for which process imputation delivers valid inference on parameters of interest.

\section*{Acknowledgments}
The authors thank Ephraim Hanks for early insights and discussions about the research. Funding for this research was provided by NOAA (RWO 103), CPW (TO 1304), and NSF (DMS 1614392). Any use of trade, firm, or product names is for descriptive purposes only and does not imply endorsement by the U.S. Government.

\bibliographystyle{fullnat}
\bibliography{two_stage_sde}

\section*{Appendix}
\appendix

\let\cleardoublepage\clearpage
\section{Simulation studies}\label{app:simulation}

\subsection{Coverage and detection regions}\label{app:coverage_detection}

To evaluate the performance of the process imputation approach, we compared the posterior distributions of $\bbeta$ to the values used to simulate the data through two summary measures. First, we computed the proportion of the time interval during which a pointwise, equal-tailed 95\% credible interval contained the true value of the function $\bbeta$, which we term the ``coverage region.'' Small coverage regions suggest that the associated method is biased or under-represents uncertainty about model parameters and may be unreliable in practice. In contrast, large coverage regions do not necessarily suggest reliability, because uncertainty about model parameters may be inflated, resulting in high coverage rates without yielding precise inference. A desirable property of posterior credible regions for model parameters is that they have high coverage, while remaining narrow enough to allow researchers to say something meaningful about relevant parameters when an effect on movement is present. We therefore computed a second summary measure defined as the proportion of the time interval during which the credible interval did not contain zero, which we term the ``detection region,'' because it represents regions for which there is substantial evidence that $\beta(t) \neq 0$. Figure 3 in the main body of the manuscript shows one example of coverage and detection regions. Our definitions of ``coverage'' and ``detection'' regions are not intended to be interpreted from a hypothesis testing perspective. Rather, they are intended to provide insight into the amount of bias and uncertainty that occur under the process imputation inferential procedure. 

\subsection{Study 1: Brownian particle in an external force field}

\subsubsection{MCMC details}

The Gaussian process AID model is specified as $\bmu^* \sim \N \lp \bzero , \bC^*\rp$,
where the covariance of $\bmu^*(t_j)$ and $\bmu^*(t_j)$ is defined as $\bC^*(t_j, t_k) = e^{-\frac{(t_j - t_k)^2}{2\phi^2}}$. The Ornstein-Uhlenbeck AID model is specified through the velocity, and was first introduced in \cite{Johnson2008}. We used the \verb=R= package \verb=crawl= \citep{Johnson2016}. The Ornstein-Uhlenbeck model has been used in several previous instances of process imputation \citep[e.g.,][]{Hanks2011, Hanks2015, Scharf2016}.

\subsubsection{Results}\label{app:results}

\begin{figure}[H]
  \centering
  \includegraphics[width = 0.67\textwidth]{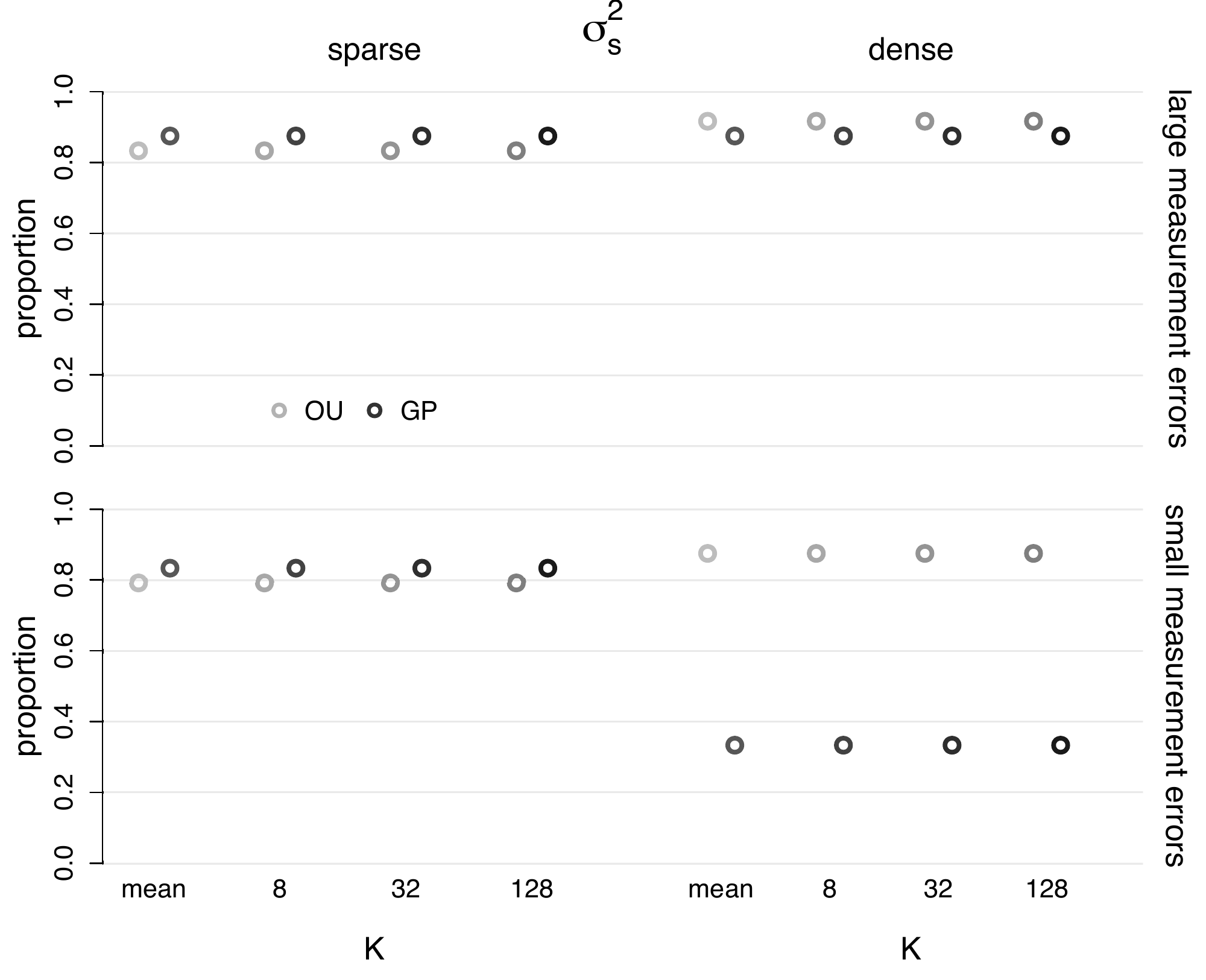}
  \caption{\footnotesize Proportion of 24 paths with 95\% credible intervals that cover the true value of $\sigma_s^2$.}
  \label{fig:sim_sigsqs_cover}
\end{figure}

Inference for the variance in the measurement error process was consistent across almost all simulation scenarios. The exception was for the case of small measurement errors and dense observation times when using the Gaussian process AID (bottom right, Figure \ref{fig:sim_sigsqs_cover}). Under this combination of circumstances, the GP AID proved too inflexible to accurately represent the true imputation distribution.

\subsubsection{Timing}
On a personal laptop with 8 GB of random access memory, and a 2.6 GHz Intel Core i5 processor, the combined two-stage fitting procedure required approximately 10 minutes of computation time per simulated data set. 

\subsection{Study 2: Comparing exact and two-stage approaches}

\subsubsection{Exact inference}\label{sec:exact_inference}

We estimated the unobserved, true location of the particle over a grid of 500 time points in addition to the observation times. The conditional distribution of $\bmu(t_i)$ is given by
\begin{align*}
  \lb \bmu(t_i) | \bmu(t_{i-1}), \beta \rb &= \N(\bmu(t_{i-1}) - \nabla \bH(\bmu(t), \beta) \Delta t_i, \Delta t_i\bI_2) \\
  &= \N(\bmu(t_{i-1}) + \beta \frac{\bc - \bmu(t_{i-1})}{\|\bc - \bmu(t_{i-1})\|_2} \Delta t_i, \Delta t_i\bI_2)
\end{align*}
where $\Delta t_i \equiv t_i - t_{i - 1}$. The conditional distributions for the observed data are given by
\begin{align*}
  \lb \bs(t_i) | \bmu(t_i), \sigma_s^2 \rb = \N \lp \bmu(t_i), \sigma_s^2 \bI_2 \rp.
\end{align*}

\subsubsection*{Priors/Hyperparameters}
\begin{align*}
  \lb \beta | \sigma_\beta^2 \rb &= \N(0, \sigma_\beta^2 = 10^5) \\
  \lb \sigma_s^2 | a_s, b_s \rb &= \IG(a_s = 10^{-3}, b_s = 10^{-4}) \\
  \sigma_0^2 &= 10^2
\end{align*}

\subsubsection*{Full Conditionals}

\paragraph{}
\textit{True path:}
\begin{align*}
  \lb \bmu(t_1 = 0) | \cdot \rb &\propto \lb \bmu(0) | \sigma_0^2 \rb
  \lb \bs(0) | \bmu(0), \sigma_s^2 \rb \\
  \begin{split}
    \lb \bmu(t_i) | \cdot \rb &\propto
    \lb \bmu(t_{i + 1}) | \bmu(t_i), \beta \rb \lb \bmu(t_i) | \bmu(t_{i-1}) \rb \\
    &\qquad \qquad \times \lb \bs(t_i) | \bmu(t_i), \sigma_s^2 \rb^{\bone{\text{observation occurred at } t_{i}}}
  \end{split} \quad 1 < i < m  \\
  \lb \bmu(t_m) | \cdot \rb &\propto  \lb \bmu(t_m) | \bmu(t_{m-1}) \rb
  \lb \bs(t_m) | \bmu(t_m), \sigma_s^2 \rb
\end{align*}

\textit{Strength of attraction:}
\begin{align*}
  \lb \beta | \cdot \rb &\propto \lb \beta | \sigma_\beta^2 \rb
  \prod_{i=2}^m \lb \bmu(t_i)|\bmu(t_{i-1}), \beta \rb
\end{align*}

\textit{Measurement error variance:}
\begin{align*}
  \lb \sigma_s^2 | \cdot \rb &\propto \lb \sigma_s^2 | a_s, b_s \rb
  \prod_{\bs(t_i)} \lb \bs(t_i) | \bmu(t_i), \sigma_s^2 \rb
\end{align*}

\subsubsection{Timing}
On a personal laptop with 8 GB of random access memory, and a 2.6 GHz Intel Core i5 processor, the combined two-stage fitting procedure required approximately 12 seconds of computation time per simulated data set, while the exact-inferential procedure required approximately 10 minutes of computation time. Caution should be exercised in generalizing the relative computational costs of the two inferential procedures explored here to other process imputation applications.

\subsubsection{Sample paths}\label{app:sample_paths}
\begin{figure}[H]
  \centering
  \includegraphics[width = 0.48\linewidth]{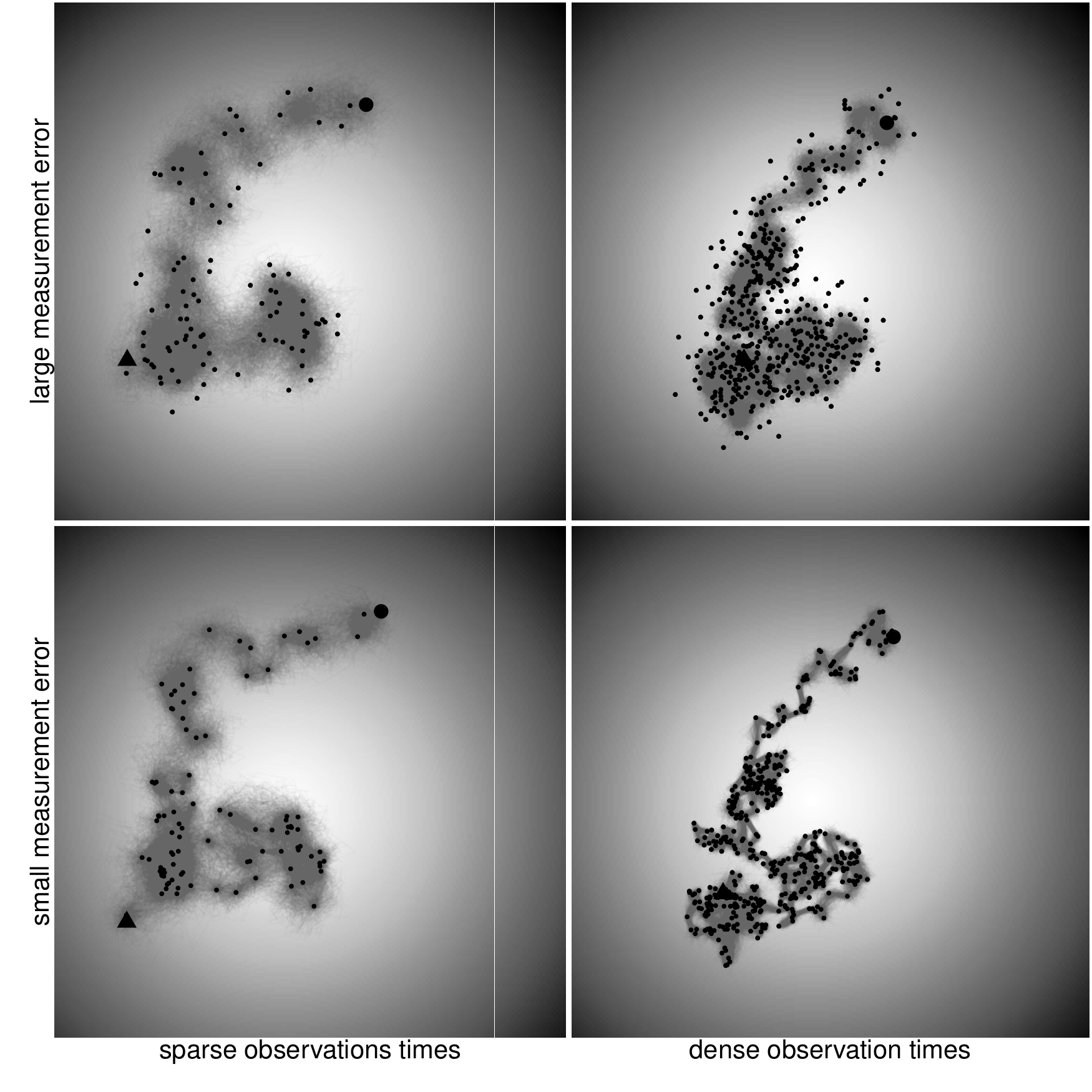}
  \caption{\footnotesize Examples of simulated data from the four different combinations of measurement error severity and observation frequency. Points represent the simulated data, gray lines represent samples from the AID, and the large circle and triangle represent the first and last observations, respectively. The background represents the potential surface, $H$, with darker shades corresponding to low values, and lighter shades corresponding to high values.}
  \label{fig:brill_eg}
\end{figure}

\section{Application}

Inference is based on a chain of $10^4$ iterations, with the first half discarded as burn-in. Evidence of convergence was determined by visual inspection of trace plots, and the Gelman-Rubin diagnostic \citep{Gelman2014}. We fit the model using a grid of values ($10^{i/4}$ for $i \in \lbr -16, \dots, 12 \rbr$) for the hyper-parameter $\sigma^2_\alpha$, and found that $10^{-1.75}$ provided an optimal amount of tuning as measured by DIC. As an alternative to score-based selection of the tuning parameter, one could instead specify a prior distribution on $\sigma^2_\alpha$. Our approach has the benefit of regularizing $\beta(t)$, which can provide improved estimation \citep[e.g.,][]{Hastie2009}.

\begin{table}[H]
  \centering
  \begin{tabular}{r|cc|c}
    \multicolumn{1}{c}{} & \multicolumn{2}{c}{posterior} & prior \\
    \hline
    parameter    & median & (2.5\%, 97.5\%) & density \\
    \hline
     $\sigma_s$ (kms) & 1.54 & (1.50, 1.58) & $\Unif(0, \infty)$     \\
     $\sigma_v$ (hour$^{-1}$) & 0.269 & (0.262, 0.275) & $\Unif(0, \infty)$     \\
     \hline
  \end{tabular}
  \caption{\footnotesize Posterior medians, 95\% credible intervals, and prior distributions for the application to northern fur seal movement.}
  \label{tab:application_results}
\end{table}

\subsection{Timing}
On a personal laptop with 8 GB of random access memory, and a 2.6 GHz Intel Core i5 processor, the combined two-stage fitting procedure required approximately 80 minutes of computation time. 

\section{Application vignette} 

This vignette shows how the two-stage process imputation procedure was implemented for the application to the movement of a Norther fur seal. 

\vspace{1em}

\url{http://www.stat.colostate.edu/~scharfh/supplemental_multiple_imputation_JABES/}

\end{document}